**Applying the Field Emission Orthodoxy Test to Murphy-Good Plots**

Mohammad M. Allaham[1], Richard G. Forbes[2] and Marwan S. Mousa[1*]

1: Surface Physics and Materials Technology lab., Department of Physics, Mutah University, Al-Karak 61710, Jordan.

2: Advanced Technology Institute & Department of Electrical and Electronic Eng., Faculty of Eng. & Physical Sciences, University of Surrey, Guildford, Surrey GU2 7XH, UK.

*mmousa@mutah.edu.jo

**Abstract.** In field electron emission (FE) studies, it is important to check and analyse the quality and validity of results experimentally obtained from samples, using suitably plotted current-voltage $[I_m(V_m)]$ measurements. For the traditional plotting method, the Fowler-Nordheim (FN) plot, there exists a so-called "orthodoxy test" that can be applied to the FN plot, in order to check whether the FE device/system generating the results is "ideal". If it is not ideal, then emitter characterization parameters deduced from the FN plot are likely to be spurious. A new form of FE $I_m(V_m)$ data plot, the so-called "Murphy-Good (MG) plot" has recently been introduced (R.G. Forbes, Roy. Soc. open sci. 6 (2019) 190912. This aims to improve the precision with which characterization-parameter values (particularly values of formal emission area) can be extracted from FE $I_m(V_m)$ data. The present paper compares this new plotting form with the older FN and Millikan-Lauritsen (ML) forms, and makes an independent assessment of the consistency with which slope (and hence scaled-field) estimates can be extracted from a MG plot. It is shown that, by using a revised formula for the extraction of scaled-field values, the existing orthodoxy test can be applied to Murphy-Good plots. The development is reported of a prototype web tool that can apply the orthodoxy test to all three forms of FE data plot (ML, MG and FN).

# 1. Introduction

The process of field electron emission (FE) occurs in very many technological contexts. This paper is about the analysis of measured current-voltage [$I_m(V_m)$] data that relate to FE processes, devices, and systems. The conventional methods of analysing this data are to make either a Fowler-Nordheim (FN) plot, i.e., a plot of the form $\log_{e,10}\{I_m/V_m^2\}$ vs $1/V_m$, or (in older work) a Millikan-Lauritsen plot, i.e., a plot of the form $\log_{e,10}\{I_m\}$ vs $1/V_m$. Such plots are often approximately straight. Emitter characterization parameters are then extracted from the slope of the plot, and from the intercept that a fitted straight line makes with the $1/V_m = 0$ axis.

With logarithmic expressions, such as $\log\{x\}$ or $\ln\{x\}$, an international convention [1] is used in this paper that the symbol $\{x\}$ means "the numerical value of $x$, when is $x$ is measured in the stated units". In this paper, in all equations, all figures and all tables, voltages are *always* measured in volts and currents are *always* measured in amperes. Brackets not part of logarithmic expressions are used normally.

A so-called *ideal* FE device/system is one where the measured $I_m(V_m)$ characteristics are determined only by unchanging system geometry and surface properties, and by the electron emission process. An ideal system is termed *orthodox* if, in addition, it is an adequate approximation to assume that emission takes place through a Schottky-Nordeim (SN) ("planar image rounded") potential-energy barrier, and hence that Murphy-Good FE theory applies.

For an orthodox device/system, FN plot (or ML plot) analysis leads to correct values for emitter characterization parameters. However, for a variety of reasons (for example, series resistance in the measurement circuit) many real FE devices/systems are not ideal (and hence cannot be orthodox). When conventional FN plot (or ML plot) analysis techniques are applied to $I_m(V_m)$ data taken from non-ideal devices/systems, then spurious values can be (and often are) derived for emitter characterization parameters.

The so-called *Orthodoxy Test* [2] is a test that can be applied to a FN plot (or to an ML plot), in order to establish whether or not the plotted data are derived from an orthodox FE device/system (and hence whether extracted characterization parameters are valid or spurious).

For a field emitter with local work function $\phi$, subject to a local electrostatic field of magnitude $F_L$, a corresponding *scaled field* $f_L$ can be defined by

$$f_L \equiv c_S^2 \phi^{-2} F_L \equiv (e^3/4\pi\varepsilon_0)\phi^{-2} F_L \equiv F_L/F_R . \tag{1}$$

where $c_S$ is the Schottky constant, $e$ is the elementary positive charge, $\varepsilon_0$ is the vacuum electric permittivity, and $F_R \equiv c_S^{-2}\phi^2$ is the reference field needed to pull the top of a SN barrier of zero-field height $\phi$ down to the Fermi level. Emitter behaviour can be described in terms of characteristic field ($F_C$) and scaled-field ($f_C$) values, usually taken as the values at the emitter apex (for a pointed emitter), or at the apex of a prominent individual emitter (for large-area field electron emitters).

The range of $f_C$-values in which emitters normally operate is well established (see spreadsheet associated with Ref. [2]). When $I_m(V_m)$ data are plotted in the form of a Fowler-Nordheim (FN) plot or a Millikan-Lauritsen (ML) plot, the plot can be used to extract values of $f_C$ that correspond to the range of electrostatic fields *apparently* used in the experiments. The Orthodoxy Test compares these apparent $f_C$-values with the known $f_C$-values at which emitters normally operate, and draws appropriate conclusions. For example, if an extracted $f_C$-value is higher than the known $f_C$-value at which an emitter melts or self-destructs, it is concluded that the FE device/system is not ideal and that any characterization results derived from the FN (or ML) plot are likely to be spurious.

Fowler-Nordheim plots came into use because the FE equation derived by FN in 1928 [3] predicted that a FN plot would be linear. But, in 1953, Burgess, Kroemer and Houston (BKH) [4] found a physical mistake in FN's thinking, and also a mathematical mistake in a related paper by Nordheim [5]. In 1956, Murphy and Good (MG) used the BKH results to develop a revised FE equation [6]. (For a modern derivation, using the International System of Quantities (ISQ), see [7].)

Murphy-Good plots are a new form of FE $I_m(V_m)$ data plot that has recently been developed [8]. They are based on improved mathematical understanding of MG theory, developed from 2006 onwards, and have the form $\log_{e,10}\{I_m/V_m^\kappa\}$ vs $1/V_m$, where the voltage exponent $\kappa$ for the SN barrier used in MG theory has the value

$$\kappa = 2 - \eta/6 . \qquad (2)$$

The parameter $\eta$ depends only on the assumed work-function $\phi$ and is given by [2,9]

$$\eta \cong 9.836239 \, (eV/\phi)^{1/2} \qquad (3)$$

More generally, in the expression $\ln\{I_m/V_m^k\}$, the value to be allocated to the general voltage exponent $k$ depends on the plot type, as shown in Table 1.

TABLE 1. Values of the voltage exponent $k$ for the three plot types under discussion

| Plot type | Voltage exponent ($k$) |
|---|---|
| Murphy-Good (MG) | $(2 - \eta/6)$ |
| Fowler-Nordheim (FN) | 2 |
| Millikan-Lauritsen (ML) | 0 |

In the work reported in this paper, we will confirm that using a MG plot is an improved method to analyse FE data. This is because modern Murphy-Good theory predicts that (unlike an ML plot or a FN plot) an MG plot will be "almost exactly" a straight line. This means that, for ideal measured current-voltage data, we can extract well-defined emitter characterisation parameters more precisely and more easily than with the older plot forms.

We also develop a form of orthodoxy test that applies to Murphy-Good plots, and report on the development of a software tool that implements this test for any of the three plot forms shown in Table 1.

## 2. Theoretical Background

The so-called *Extended Murphy Good (EMG) equation* [8] for the local emission current density (LECD) $J_L^{EMG}$ can be written in the linked form

$$J_L^{EMG} = \lambda J_{kL}^{SN} \tag{4}$$

$$J_{kL}^{SN} \equiv a\phi^{-1}F_L^2 \exp[-v_F b\phi^{3/2}/F_L] \ . \tag{5}$$

Here: $\lambda$ is an "uncertainty factor", of unknown functional dependence or value, called the *local pre-exponential correction factor*; $J_{kL}^{SN}$ is called the *kernel LECD for the SN barrier*; and *a* and *b* are the *first* and *second FN constants* as usually defined ([2], also see Table 2 here). The parameter $v_F$, which acts as the barrier-form correction factor for the SN barrier defined by $\phi$ and $F_L$, is a particular value of a special mathematical function $v(x)$, where $x$ is the Gauss variable (i.e., the independent variable in the Gauss Hypergeometric Differential Equation), and usually is adequately given by the simple good approximation

$$v_F = v(x=f) \approx 1 - f + (f/6)\ln f \ , \tag{6}$$

where $f$ is to be interpreted as the *local scaled field* $f_L$ as defined above, or as its characteristic value $f_C$.

Integrating Eq. (4) over the emitter surface, we can find the *total emission current* $I_e^{EMG}$ as

$$I_e^{EMG} = A_n^{EMG} \lambda J_{kL}^{SN}, \tag{7}$$

where $A_n^{EMG}$ is the related *notional emission area*. On defining the *formal emission area (for the SN barrier)* by $A_f^{SN}$ by

$$A_f^{SN} = \lambda A_n^{EMG}, \tag{8}$$

and on assuming that the measured current $I_m$ is equal to $I_e^{EMG}$, we obtain the $I_m(F_C)$ form of the EMG equation, as

$$I_m(F_C) = A_f^{SN} a \phi^{-1} F_C^2 \exp[-v_F b \phi^{3/2} / F_C]. \tag{9}$$

The characteristic barrier field $F_C$ can be related to the measured voltage $V_m$ by the formula

$$F_C = V_m / \zeta_C, \tag{10}$$

where $\zeta_C$ is a system-geometry parameter called the *characteristic voltage conversion length* (*VCL*). This parameter $\zeta_C$ is constant for ideal FE devices and systems, because their $I_m(V_m)$ characteristics are determined only by the emission process and unchanging system geometry and surface properties. Additional background theory is given in [7-9].

## 3. The theory of Murphy-Good (MG) plots

### *3.1 Development of a Theoretical Equation for the MG Plot*

This Section reviews the theory [8] of Murphy-Good plots. To develop the theory, it is necessary to put Eq. (11) into so-called *scaled form*. From Eq. (1), for characteristic values, we have $F_C \equiv c_S^{-2} \phi^2 f_C$. We can define two scaling parameters $\eta(\phi)$ and $\theta(\phi)$ by

$$\eta(\phi) = b c_S^2 \phi^{-1/2} , \tag{11}$$

$$\theta(\phi) = a c_S^{-4} \phi^3 . \tag{12}$$

Inserting these three relationships into Eq. (9) yields the *scaled EMG equation*

$$I_m(f_C) = A_f^{SN} \theta f_C^2 \exp[-v_F \eta / f_C], \tag{13}$$

where, for simplicity, we do not explicitly show that $\eta$ and $\theta$ depend on $\phi$. Making use of approximation (6), with $f=f_C$, yields (after some algebraic manipulation)

$$I_{m(}(f_C) = A_f^{SN} \theta f_C^{(2-\eta/6)} \exp[\eta] \exp[-\eta/f_C]. \tag{14}$$

Equation (13) now needs to be converted into a form where the measured voltage $V_m$ is the independent variable. For an ideal device/system, the *reference measured voltage* $V_{mR}$ is related to the *reference field* $F_R$ (that corresponds to $f_C=1$) by

$$V_{mR} \equiv F_R \zeta_C. \tag{15}$$

Combining this with Eq. (10) and definition (1) for characteristic scaled field $f_C$ yields

$$f_C = F_C/F_R = V_m/V_{mR}. \tag{16}$$

In Eq. (14), we use Eq. (2) to replace $(2-\eta/6)$ by $\kappa$, and use Eq. (16) to replace $f_C$, yielding

$$I_m(V_m) = \{A_f^{SN} \theta \exp[\eta] V_{mR}^{-\kappa}\} V_m^{\kappa} \exp[-\eta V_{mR}/V_m]. \tag{17}$$

Dividing both sides by $V_m^{\kappa}$, and taking the *natural logarithms* of both sides, gives us the equation for the *theoretical MG plot*:

$$\ln\{I_m/V_m^{\kappa}\} = \ln\{A_f^{SN} \theta \exp[\eta] V_{mR}^{-\kappa}\} - \eta V_{mR}/V_m. \tag{18}$$

### *3.2 Extracting parameters from a MG plot*

MG plots have the form $\log_{e,10}\{I_m/V_m^\kappa\}$ vs $1/V_m$, but we discuss only the natural logarithmic form here. We can define the following expressions:

$$Z \equiv 1/V_m \, , \tag{19}$$

$$Y \equiv \ln\{I_m/V_m^\kappa\} \, , \tag{20}$$

$$\alpha \equiv A_f^{SN} \, \theta \exp[\eta] \, V_{mR}^{-\kappa} \, , \tag{21}$$

$$\beta \equiv -\eta V_{mR} \, . \tag{22}$$

For a given work-function value, $\beta$, $\alpha$ and $\ln\{\alpha\}$ are constants. On substituting Eqs. (19) to (22) into Eq. (18), we obtain the linear equation

$$Y(X) = \ln\{\alpha\} + \beta Z \, . \tag{23}$$

Thus, $\ln\{\alpha\}$ is the *theoretical intercept* of the MG plot and $\beta$ is its *theoretical slope*. It can also be seen that

$$\alpha \, |\beta|^\kappa \equiv A_f^{SN}\theta\eta^\kappa \exp[\eta] = A_f^{SN}\theta\eta^2\eta^{-\eta/6} \exp[\eta] \, . \tag{24}$$

From eqs (11) and (12), it follows that $\theta\eta^2 = ab^2\phi^2$, so we obtain

$$\alpha |\beta|^\kappa = A_f^{SN} \theta \eta^\kappa \exp[\eta] = A_f^{SN}(ab^2\phi^2)(\eta^{-\eta/6}\exp[\eta]). \tag{25}$$

Let $S_{MG}^{fit}$ denote the slope of a straight line fitted to an experimental MG plot (made using natural logarithms), and let $\ln\{R_{MG}^{fit}\}$ denote the intercept that this line makes with the $(1/V_m)=0$ axis. It follows from the equations above that an extracted value of formal emission area $A_f^{SN}$ can be obtained from the extraction formula

$$\{A_f^{SN}\}^{extr} = \Lambda_{MG} R_{MG}^{fit} |S_{MG}^{fit}|^\kappa, \tag{26}$$

where the *extraction parameter for the MG plot*, $\Lambda_{MG}(\phi)$, is given by

$$\Lambda_{MG}(\phi) = 1/\left[(ab^2\phi^2)(\eta^{-\eta/6}\exp[\eta])\right] \tag{27}$$

Examples of the numerical dependence of $\Lambda_{MG}(\phi)$ on $\phi$ are given in [8].

It also follows that extracted values of $V_{mR}$ and $\zeta_C$ can be obtained from

$$\{V_{mR}\}^{extr} = -S_{MG}^{fit}/\eta, \tag{28}$$

$$\{\zeta_C\}^{extr} = -S_{MG}^{fit}/b\phi^{3/2}. \tag{29}$$

Jordan Journal of Physics, in press, February 2020

For large area field electron emitters (LAFEs), an extracted value of a characteristic (dimensionless) field enhancement factor (FEF) $\gamma_{MC}$, can then be obtained from

$$\{\gamma_{MC}\}^{extr} = d_M / \{\zeta_C\}^{extr}, \tag{30}$$

where $d_M$ is the macroscopic distance used to defined the FEF (often, but not necessarily, the separation between two parallel planar plates).

Once a value has been extracted for $V_{mR}$, Eq. (14) can be used to determine (for an ideal device/system) the characteristic scaled-field value $f_C$ that corresponds to any measured voltage. This is equivalent to using the extraction formula

$$\{f_C\}^{extr} = V_m / V_{mR} = -(\eta / S_{MG}^{fit}) / (1/V_m). \tag{31}$$

This can be contrasted with the formula for extracting $f_C$-values from an FN plot made against $1/V_m$, which is [2]

$$\{f_C\}^{extr} = V_m / V_{mR} = -(s_t \eta / S_{MG}^{fit}) / (1/V_m). \tag{32}$$

Clearly, the MG-plot formula does not contain the slope correction factor $s_t$.

### 4. Orthodoxy test for a Murphy-Good plot

*4.1 Description of the Test*

Since the orthodoxy test is based on comparing extracted ranges of $f_C$ with acceptable and unacceptable ranges of $f_C$, as defined in Ref. [2] and shown in Tables 2 and 3 below, it is straightforward to apply an orthodoxy test to a Murphy-Good plot, by using Eq. (31) to extract apparent $f_C$-values.

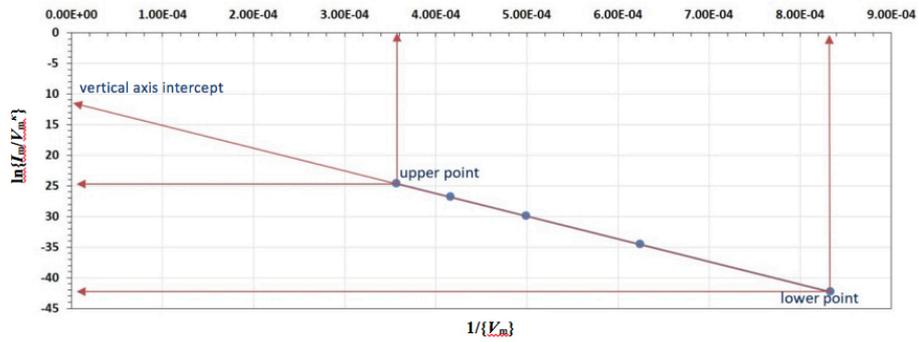

FIG. 1. Murphy-Good (MG) plot for the data sample shown in Table 5, showing the upper and lower data points that define the range of voltages used. This plot covers the range 1.2 to 2.8 kV (corresponding to the scaled-field range 0.15 to 0.35). For clarity, "computer notation" is used for the horizontal-axis numerical values. Voltages are measured in volts and currents in amperes.

To describe the test procedure, we use the simulated ideal $I_m(V_m)$ MG plot shown in Fig. 1. The method of generating this plot is described in detail in Section 4.2. The test procedure is as follows.

(1) Fit a straight line to the experimental (or simulated) plot. Regression techniques can be used, but usually defining a straight line with a ruler is good enough.

(2) Identify the position on the line that has the same *X*-coordinate ($X_{up}$) as the *lowest-X* data-point you wish to use, and determine the *Y*-coordinate ($Y_{up}$) of this position on the line. These coordinates ($X_{up}$, $Y_{up}$) define the *upper point* shown in Fig. 1. (The apparently contradictory terminology arises because "up" refers to the related value of $V_m$, rather than of $1/V_m$.)

(3) Carry out a similar procedure for the *lower point* shown in Fig. 1.

(4) Evaluate the (negative) *slope* $S_{MG}^{fit}$ of the fitted line, using the formula

$$S_{MG}^{fit} = (Y_{up} - Y_{low})/(Z_{up} - Z_{low}). \tag{33}$$

(5) Extract the range-defining scaled-field values, by applying Eq. (31) to the ($1/V_m$)-values that define the ends of the range, as follows:

$$\{f_C\}_{low}^{extr} = -(\eta/S_{MG}^{fit})/(1/V_m)_{low}, \tag{34}$$

$$\{f_C\}_{up}^{extr} = -(\eta/S_{MG}^{fit})/(1/V_m)_{up}. \tag{35}$$

(6) Apply the test condition, as derived from Tables 2 and 3.

In Table 2, $\{f_C\}_{low}^{extr}$ is the extracted $f_C$-value for the lower point, $\{f_C\}_{up}^{extr}$ is the extracted $f_C$-value for the upper point, and the superscripts (A/NA) indicate the allowed/ disallowed limits for the extracted $f_C$-values. (For simplicity, the subscript "C" is omitted in the tables.) Table 3 shows the values of these limits for various work-function values [2]. If necessary, linear interpolation between these limits can be used.

TABLE 2. General criteria for the orthodoxy test.

| Condition | Result | Explanation |
|---|---|---|
| $f_{\text{low}}^{\text{A}} \leq f_{\text{low}}^{\text{extr}}$ AND $f_{\text{up}}^{\text{extr}} \leq f_{\text{up}}^{\text{A}}$ | Pass | Reasonable range |
| $f_{\text{low}}^{\text{extr}} \leq f_{\text{low}}^{\text{NA}}$ OR $f_{\text{up}}^{\text{NA}} \leq f_{\text{up}}^{\text{extr}}$ | Fail | Clearly unreasonable range |
| $f_{\text{low}}^{\text{NA}} \leq f_{\text{low}}^{\text{extr}} \leq f_{\text{low}}^{\text{A}}$ | Inconclusive | More investigation is needed |
| $f_{\text{up}}^{\text{A}} \leq f_{\text{up}}^{\text{extr}} \leq f_{\text{up}}^{\text{NA}}$ | Inconclusive | More investigation is needed |

TABLE 3. Orthodoxy-test range-limits, as function of work function $\phi$.

| $\phi$ (eV) | $f_{\text{low}}^{\text{NA}}$ | $f_{\text{low}}^{\text{A}}$ | $f_{\text{up}}^{\text{A}}$ | $f_{\text{up}}^{\text{NA}}$ |
|---|---|---|---|---|
| 5.50 | 0.09  | 0.14 | 0.41 | 0.69 |
| 5.00 | 0.095 | 0.14 | 0.43 | 0.71 |
| 4.50 | 0.10  | 0.15 | 0.45 | 0.75 |
| 4.00 | 0.105 | 0.16 | 0.48 | 0.79 |
| 3.50 | 0.11  | 0.17 | 0.51 | 0.85 |
| 3.00 | 0.12  | 0.18 | 0.54 | 0.91 |
| 2.50 | 0.13  | 0.20 | 0.59 | 0.98 |

The physical meanings of the two "fail" ranges are easy to state. The "low-$f_C$" one corresponds to the situation where the extracted $f_C$-value is thought too low for a measurable current to be detected in normal experiments; the "upper-$f_C$" one corresponds to the situation where the extracted $f_C$-value is higher than $f_C$-values at which the emitter is known to electroform (i.e., change shape due to atomic migration) or self-destruct. In both cases, it has to be concluded that the FE device/system is not ideal, and that any characterization parameters extracted from the plot are likely to be spurious.

At a recent conference [10], we reported on the development of a prototype of a web tool that can apply an orthodoxy test to either an ML or an FN plot, and—if the test is passed—extract values of relevant emitter characterization parameters. The output of this prototype is shown in Fig. 2. During the work reported here we have extended this prototype to include MG plots. At the time of writing, this prototype can be found at link [11]. The web tool is still under development, and the final version will be made openly available in due course. It is also planned to develop a downloadable spreadsheet version

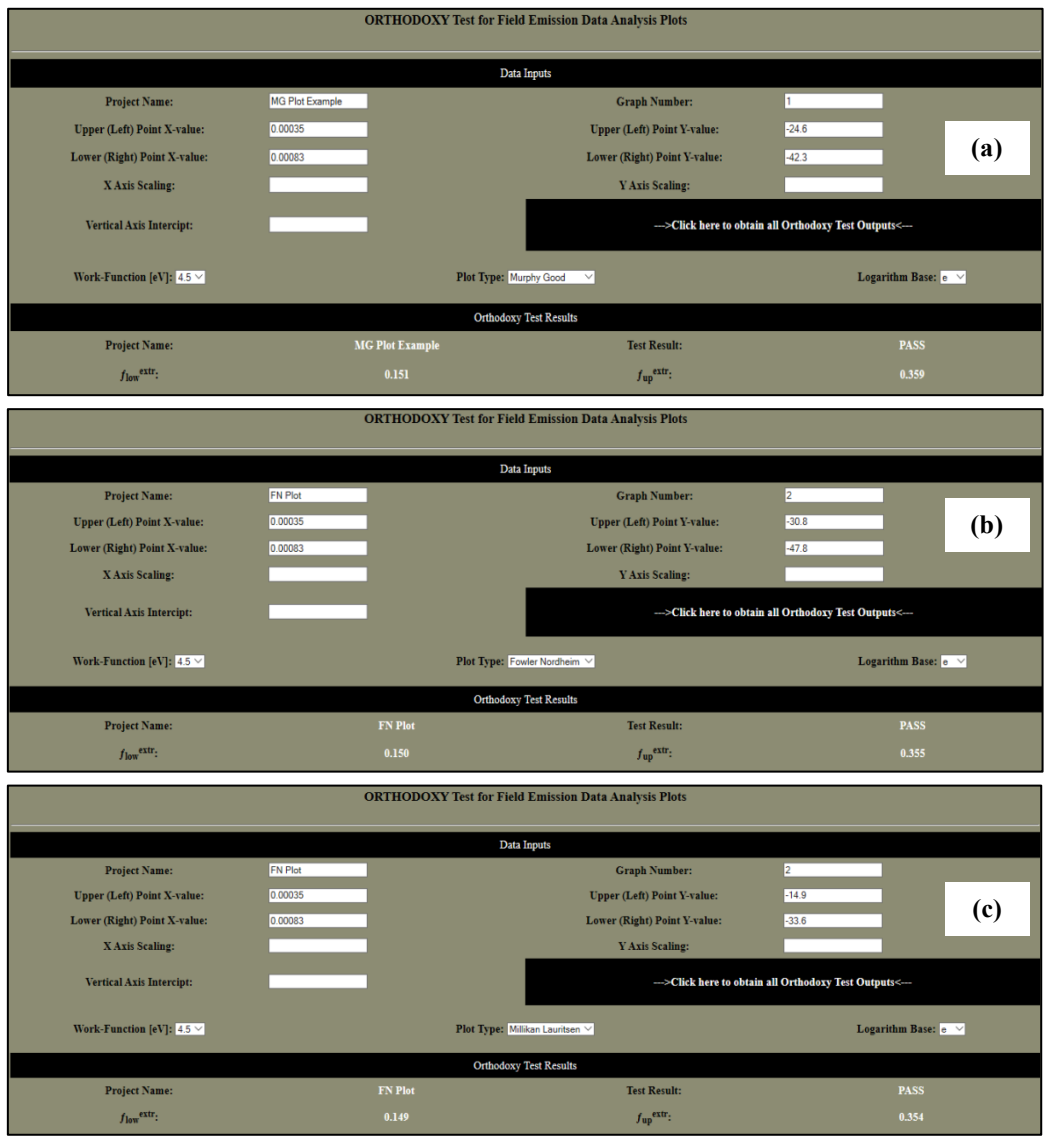

FIG. 2. Current outputs of Orthodoxy Test web tool, for: (a) MG plot; (b) FN plot; (c) ML plot.

*4.2 Consistency of scaled-field extraction for MG plot*

For an MG plot, the physical consistency of the extraction process can be checked in the following simple way. Highly precise simulated ideal $I_m(V_m)$ data-sets can be generated by using: (a) Eqs (13) and 14); (b) chosen values for system input parameters $\phi$, $V_{mR}$ and $A_f^{SN}$; (c) a high-precision (HP) formula for $v(f)$, given in [7] and also in the Appendix to [8] (in the range $0 \leq f \leq 1$, $v(f)$ varies from 1 down to 0, and the HP formula yields $v(f)$ values known to have error less than $8 \times 10^{-10}$); and (d) a chosen set of values for $f_C$.

Values of parameters used in (or related to) this simulation are shown in Table 4. The chosen values of $f_C$ (as shown in Table 5) lie in the range $0.15 \leq f_C \leq 0.35$. This range is used because it is known [2] that, for tungsten FE devices, experimental $f_C$-values often lie within this range. Table 5 also shows resulting simulated values of quantities relevant to drawing the MG plot shown in Fig. 1.

For simplicity, the "lower" and "upper" data points on the MG plot are assumed to have the horizontal ("$Z$") and vertical ("$Y$") coordinates given by the values in columns 4 and 5 of Table 5, for the $f_C$-values 0.15 and 0.35. The resulting "fitted" slope, $S_{MG}^{fit}$, derived using Eq. (33), is shown in Table 4. The corresponding extracted values $\{f_C\}^{extr}$ are shown as the last column in Table 5, and are consistent with the input values, apart from a small systematic error of 0.27%. The cause of this error is the small discrepancy between the highly precise numerical formula for $v(f)$ used in the simulations and the "simple good approximation" (6) used to develop MG-plot theory, which is known to have an error of this order of magnitude. For practical purposes the error is negligible.

Table 4 also shows extracted values of reference measured voltage and of voltage conversion length. Again, these are very close to the input values.

The various good agreements just discussed, between extracted values and input values, serve to demonstrate the physical self-consistency of the MG plot and related extraction formulae, when these are applied to an orthodox FE device/system.

TABLE 4. Parameters used for preparing simulation data for MG plot, and extracted outputs related to its slope. Universal constants are shown to seven significant figures. Other parameters are shown to four or five figure precision. Asterisks indicate the chosen input-parameter values.

| *Parameter name* | Symbol | Numerical value | Units |
|---|---|---|---|
| *Input and related data* | | | |
| First FN constant | $a$ | 1.541 434 | µA eV V$^{-2}$ |
| Second FN constant | $b$ | 6.830 890 | eV$^{-3/2}$ (V/nm) |
| Schottky constant | $c_S$ | 1.999 985 | eV (V/nm)$^{-1/2}$ |
| Local work function* | $\phi$ | 4.500 | eV |
| Reference field | $F_R$ | 14.06 | V/nm |
| Exponent scaling factor | $\eta$ | 4.637 | – |
| Pre-exponent scaling factor | $\theta$ | 6.774×10$^{13}$ | A m$^{-2}$ |
| Voltage exponent (SN barrier) | $\kappa$ | 1.227 | – |
| Reference measured-voltage* | $V_{mR}$ | 8000 | V |
| Voltage conversion length | $\zeta_C$ | 56.90 | Nm |
| Formal area (SN barrier)* | $A_f^{SN}$ | 100.0 | Nm |
| *Extracted data* | | | |
| Fitted slope | $S_{MG}^{fit}$ | –3.6995×10$^4$ | Np V |
| Extracted value of $V_{mR}$ | $\{V_{mR}\}^{extr}$ | 7987 | V |
| Extracted value of $\zeta_C$ | $\{\zeta_C\}^{extr}$ | 56.73 | Nm |

TABLE 5. Typical simulation data for a current-voltage [$I_m(V_m)$] based MG Plot, and related extracted values of characteristic scaled field $f_C$.

| $f_C$ | $V_m$ | $I_m$ | $1/V_m$ | $\ln\{I_m/V_m^\kappa\}$ | $\{f_C\}^{extr}$ | % error |
|---|---|---|---|---|---|---|
| – | (V) | (A) | (V$^{-1}$) | – | – | |
| 0.15 | 1200 | 2.755×10$^{-15}$ | 8.33×10$^{-4}$ | –42.2 | 0.15040 | 0.27% |
| 0.20 | 1600 | 8.747×10$^{-12}$ | 6.25×10$^{-4}$ | –34.5 | 0.20054 | 0.27% |
| 0.25 | 2000 | 1.172×10$^{-9}$ | 5.00×10$^{-4}$ | –29.9 | 0.25067 | 0.27% |
| 0.30 | 2400 | 3.196×10$^{-8}$ | 4.17×10$^{-4}$ | –26.8 | 0.30081 | 0.27% |
| 0.35 | 2800 | 3.488×10$^{-7}$ | 3.57×10$^{-4}$ | –24.6 | 0.35094 | 0.27% |

A more detailed investigation was also carried out, in order to confirm the high consistency level with which characterization parameters can be extracted from a MG plot for an orthodoxly behaving FE device/system (if likely statistical errors in measured data are disregarded). This investigation was similar to, but performed independently of, that described in [8].

A data-set was created by using the same basic input data as above, but using $f_C$-values at intervals of 0.01 in the range $0.15 \leq f_C \leq 0.35$. This $f_C$-range was then divided into four different smaller ranges, in order to check how much the extracted slope $S_{MG}^{fit}$ varied, depending on the particular range chosen. As shown in the first five columns of Table 6, it was found that, to a very good level of precision, the extracted MG-plot slope does not depend on the range used.

TABLE 6. Simulation of extraction of fitted slope $S_{MG}^{fit}$, for a MG plot, and fitted slope $S_{FN}^{fit}$ for a FN plot, for various different ranges of characteristic scaled field $f_C$ (and hence of predicted measured voltage $V_m$). Input data as in Table 4.

| $(f_C)_{low}$ | $(f_C)_{up}$ | $(V_m)_{low}$ (V) | $(V_m)_{up}$ (V) | $S_{MG}^{fit}$ (Np V) | $S_{FN}^{fit}$ (Np V) |
|---|---|---|---|---|---|
| 0.16 | 0.20 | 1280 | 1600 | −37006 | −35902 |
| 0.21 | 0.25 | 1680 | 2000 | −36992 | −35577 |
| 0.26 | 0.30 | 2080 | 2400 | −36981 | −35256 |
| 0.31 | 0.35 | 2480 | 2800 | −36974 | −34938 |
|  |  |  | \|Total variation\|: | 32 | 964 |

For the MG plot, the residual variations shown in Table 6 presumably result, as before, because the simple good approximation (6), used to derive MG-plot theory, is not an exactly precise expression for $v(f)$.

## 5. Comparisons between the plot types

It is known that a Fowler-Nordheim plot is not expected to be an exactly straight line, and that this gives rise to difficulties when interpreting a straight line fitted to an experimental FN plot. Within the framework of the prevailing "smooth planar metal emitter (SPME)" methodology almost-universally used for interpreting FE current-voltage data (see Section 6), the Murphy-Good plot was designed [8] to be "very nearly" a straight line, and to remove these interpretation difficulties.

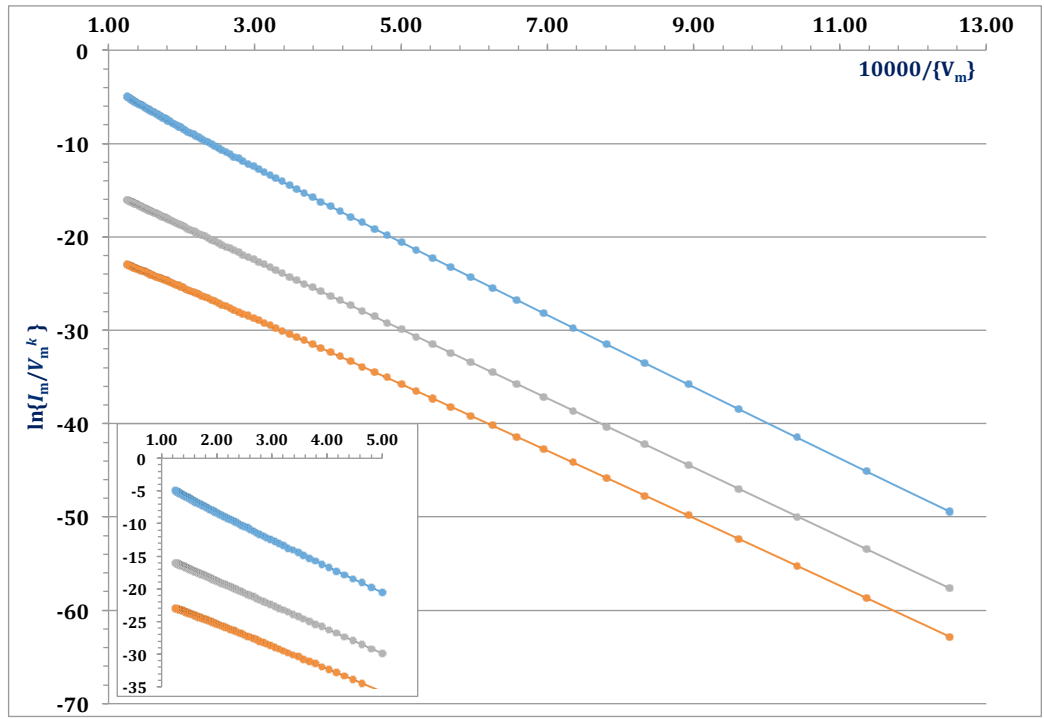

FIG. 3. Comparison between plot types. The ML plot (top) has voltage exponent $k=0$; MG plot (middle) has $k=\kappa$; FN plot (bottom) has $k=2$. Voltages are measured in volts and currents in amperes.

The superior quality of the MG plot is illustrated in Fig. 3, which shows the same set of basic $I_m(V_m)$ data plotted in the three ways under discussion. These plots are each based on a large set of data points, distributed over a wider $f_C$-value range than was used to produce Fig. 1. The MG plot is "very nearly" straight, as $1/V_m$ gets smaller, whereas the FN plot curves slightly downwards and the ML plot curves upwards. The differences are small and normally difficult to see. However,—because the "lever effect" operates when extracting, from experimental data points well away from the $(1/V_m)=0$ axis, a value for the intercept on the axis—these small differences significantly affect the accuracy with which the intercept value can be extracted.

To illustrate this quantitatively, we carried out for a FN plot a simulation exercise analogous to that described in Section 4.2 for an MG plot, using the same set of values of $f_C$, $V_m$ and $I_m$. The results for the FN-plot fitted slope $S_{FN}^{fit}$ are listed in the last column of Table 6: the variation in $S_{FN}^{fit}$ (around 3%) is around 30 times higher than the variation in the fitted slope $S_{MG}^{fit}$ for the MG plot. This again demonstrates the superiority of the MG plot as a tool for the precise analysis of $I_m(V_m)$ data (within the framework of SPME methodology).

Obviously, this difference between MG plots and FN plots aligns with the fact that Eq. (30) contains the slope correction factor $s_t$ but Eq. (29) does not. It is known (e.g. [7]) that the slope correction function $s(f)$ is a function of $f$, albeit a weak one. Thus, the "fitting value" $s_t$ [$\equiv s(f_t)$] will be a function of the fitting value $f_t$ ($f_t$ is the value of characteristic scaled-field at which the experimental plot and the tangent to the theoretical plot are assumed to be parallel.) However, normal practice is to take $s_t$ as having the constant value 0.95; this is part of the cause of the observed discrepancy. (One of the problems with precise FN-plot analysis is determining the precise value of $s_t$).

Note that the numerics presented here have been generated specifically for the purpose of confirming the theoretical performance of a MG plot, and comparing this with that of a FN plot. These numerics should **not** be interpreted as likely errors when MG plots are used to interpret real experimental data. In the application of FN plots and MG plots to real experimental data, other factors come into play, including noise in the data, non-ideality, and weaknesses in SPME methodology.

## 6. Applicability to non-metals and non-planar emitters

Obviously, the whole discussion here has been within the framework of the near-universal experimentalists' assumption (whatever the material they are working with) that, theoretically, emitters can be treated as if: (a) the existence of atomic structure can be disregarded; (b) emission is coming from a limited area of a smooth planar surface of very large extent; and (c) Sommerfeld free electron-metal theory applies. This has been called elsewhere [8] the "smooth planar metal emitter" (SPME) methodology. The above set of assumptions is obviously wildly unrealistic for some modern field electron emitters (in particular, for low-apex-radius carbon nanotubes). Clearly, for such emitters important questions are: "how should FE $I_m(V_m)$ data-analysis techniques be modified?" and (in the context of the present paper) "do new test(s) of FE 'ideality' need to be developed to replace the orthodoxy test?"

To a large extent, these questions are outside the scope of the present paper. The MG plot is a method for improving the precision of data analysis within the framework of SPME methodology, and the focus of this paper has been on developing and testing a form of orthodoxy test applicable to a MG plot. These things have merit in themselves. Nevertheless, the following points deserve making.

In reality, the situation is more favourable than it might seem at first sight. (1) (Except perhaps for *very* sharp emitters) most of the causes of non-orthodoxy in FE devices/systems are associated with breakdown of the assumption that $V_{mR}$ is constant, rather than with breakdown of the assumption that the emission is adequately described by Murphy-Good theory. But the causes of breakdown in the constant-$V_{mR}$ assumption are much the same, whatever the emitter material. (2) In breakdown of the assumption that the EMG FE equation applies, the biggest factor will probably be that the true barrier is not a SN barrier. But, except possibly for *very* sharply curved emitters, the effect on FN-plot analysis is known to be a change in the value of $s_t$ by a relatively small amount (typically of the order of a few percent); equivalently, one would expect (qualitatively) that the numerical validity of MG-plot analysis would be affected only in a small way. (3) Differences in the electron supply function would be expected to have only a very small effect on the FN plot or MG plot slope, since they primarily affect the plot intercept. (4) With very sharp emitters it is likely that both FN and MG plots would be noticeably curved, and thus obviously non-orthodox. (5) Lack of strict applicability of orthodoxy test theory is more likely to result in a false determination that the FE device-system is non-orthodox, than a false determination that the FE device/system is orthodox. In fact, for further development of FE theory, a false determination of the first kind is of limited importance, because only results that pass the orthodoxy test are of scientific use. (6) The orthodoxy test is an "engineering triage" test, with generous margins of uncertainty built in.

Although more-exact tests for "ideality/non-ideality" may be developed in due course, this seems unlikely to happen soon. For the time being, we consider that the orthodoxy test, whether applied to FN or MG plots, is a technological test of "ideality or otherwise" that is sufficient for purpose.

A final point is that development of data-analysis theory for non-metal emitters and for very sharp emitters is inhibited by lack of sufficiently good understanding of emission theory for such emitters. Probably a higher strategic priority is to first develop a form of data-analysis theory that deals with metal emitters that have the shape of pointed needles or of rounded posts, or are otherwise sharp (but not "very sharp"). This topic is beginning to be an active area of research, and some of the issues involved have recently been discussed (see doi:10.13140/RG.2.2.32112.81927 and doi:10.13140/RG.2.2.35337.19041).

### 7. Summary

This paper has reviewed the theory of Murphy-Good (MG) plots, and has made some comparisons with the theories of Fowler-Nordheim (FN) plots and Millikan-Lauritsen (ML) plots. If experimental current-voltage characteristics conform to the so-called "Extended MG equation", then an experimental MG plot is expected to be "very nearly" a straight line, and "much more nearly straight" than either an FN plot or a ML plot. We have confirmed this to be the case. Also, by extracting (from a simulated MG plot) slope values that correspond to different voltage ranges, we have shown that (for an MG plot taken from a orthodoxly behaving FE device/system) the process of slope extraction generates very consistent slope values, irrespective of voltage range

Because values of characteristic scaled field $f_C$ can be extracted from a MG plot by using Eq. (31), an orthodoxy test can be applied to a MG plot by using the same rules about scaled-field ranges as apply in the case of the FN plot. A prototype web tool that can apply the orthodoxy test to any of the three types of FE data plot discussed (ML, MG or FN) is partially completed, though still under development. It is planned that there should eventually also be a downloadable spreadsheet version.

A particular application that we have in mind is to use the new data-analysis techniques discussed here, namely Murphy-Good plots and the related orthodoxy test, to enhance the analysis of experimental results obtained in the field emission laboratory at Mu'tah University.